\documentclass[showpacs,prl,superscriptaddress,nofootinbib, twocolumn]{revtex4}
\usepackage{graphicx}
\usepackage{color}
\usepackage{amssymb,amsfonts,amsmath}
\usepackage{natbib}
\usepackage{epstopdf}
\usepackage{subfigure}

\begin{document}

\title{Intertangled stochastic motifs in networks of excitatory-inhibitory units}

\author{Clement Zankoc} 
\affiliation{Dipartimento di Fisica e Astronomia and CSDC, Universit\`{a} degli Studi di Firenze, via G. Sansone 1, 50019 Sesto Fiorentino, Italia}
\affiliation{INFN Sezione di Firenze, via G. Sansone 1, 50019 Sesto Fiorentino, Italia}
\author{Duccio Fanelli}  \affiliation{Dipartimento di Fisica e Astronomia and CSDC, Universit\`{a} degli Studi di Firenze, via G. Sansone 1, 50019 Sesto Fiorentino, Italia}
\affiliation{INFN Sezione di Firenze, via G. Sansone 1, 50019 Sesto Fiorentino, Italia}
\author{Francesco Ginelli} 
\affiliation{SUPA, Institute for Complex Systems and Mathematical Biology, Kings College, University of Aberdeen, Aberdeen AB24 3UE, United Kingdom}
\author{Roberto Livi} 
\affiliation{Dipartimento di Fisica e Astronomia and CSDC, Universit\`{a} degli Studi di Firenze, via G. Sansone 1, 50019 Sesto Fiorentino, Italia}
\affiliation{INFN Sezione di Firenze, via G. Sansone 1, 50019 Sesto Fiorentino, Italia}

\begin{abstract}
A stochastic model of excitatory and inhibitory interactions which bears universality traits is introduced and studied. The endogenous component of noise, stemming from finite size corrections, 
drives robust inter-nodes correlations, that persist at large large distances. Anti-phase synchrony at small frequencies is resolved on adjacent nodes and found to promote the spontaneous generation of long-ranged stochastic patterns, that invade the network as a whole. These patterns are lacking under the idealized deterministic scenario, and could provide novel hints on how living systems implement and handle a large gallery of delicate computational tasks. 
\end{abstract}

\pacs{02.50.Ey,05.40.-a, 87.18.Sn, 87.18.Tt} 

\maketitle

Living systems execute an extraordinary plethora of complex functions, that result from the intertwined interactions among key microscopic actors~\cite{alberts}. 
Positive and negative feedbacks appear to orchestrate the necessary degree of macroscopic coordination~\cite{pikovsky}, by propagating information to distant sites while 
supporting the processing steps that underly categorization and decision-making. Excitatory and inhibitory circuits play, in this respect, a role of 
paramount importance.  As an example, networks of excitatory and inhibitory neurons constitute the primary computational units in the brain cortex~\cite{kandel, WC1, WC2}.  
They can flexibly adjust to different computational modalities,  as triggered by distinct external stimuli \cite{gollo, nicosia}.  
Genetic regulation is also relying on sophisticated inhibitory and excitatory loops~\cite{albert, hasty, elowitz, isaacs}. Specific genes are customarily assigned to the nodes of a given constellation, 
which can be abstractly pictured as a complex network~\cite{caldarelli, boccaletti, boccaletti2}. Weighted edges between adjacent nodes encode for the characteristics of the interaction. 
Simple deterministic models can be put forward to reproduce at the mesoscopic, coarse grained level, the prototypical evolution of excitatory and inhibitory units, organized in two mutually competing populations.
Continuous variables are customarily introduced to quantify the activity of each selected population. Non linearities prove crucial to adequately represent the threshold mechanism of 
activation that modulates, from neurons to genes, the system response~\cite{murray, goldbeter}. One can then assemble large networks with designated topology,
by replicating the aforementioned module on each node of the collection and incorporating the specific nature of the coupling \cite{kouvaris}. Stationary patterns \cite{kouvaris2} of asynchronous activity 
for the scrutinized species can eventually emerge, following symmetry breaking instabilities that necessitate a fine tuning of the parameters involved. 
These patterns could define the basic architectural units for natural systems to perform efficient computations~\cite{sole, sole1, sole2}. 

As opposed to the deterministic formulation, an individual-based description -- hence intrinsically stochastic for any finite population -- can be invoked~\cite{gardiner, vankampen}. This amounts to characterizing the microscopic dynamics via transition rates, that govern the interactions among individuals and with the surrounding environment. The stochastic component ultimately results from the inherent discreteness of the system, and can 
significantly modify the idealized mean-field predictions. Endogenous fluctuations induced by the finiteness of the system can, for instance, seed the emergence of regular oscillations, when parameters are set so to drive deterministic convergence towards a trivial equilibrium~\cite{bartlett, mckanenewman, dauxois, asslani, asslani2, biancalani, woolley, bressloff}. 

In this Letter, we put forward a minimal model for discrete collections of 
excitatory and inhibitory agents in mutual interaction with excitatory and inhibitory loops, bearing universality traits 
in light of its inherent simplicity. Endogenous-noise induces quasi-cyclic dynamics that display unusual long range correlations, persisting over arbitrary large network structures. 
In particular, we find that in a large parameter region the level of activity associated to homologous species hosted on 
contiguous nodes exhibits anti-phase synchrony and that dynamical patterns of anti-phase locked stochastic oscillators sets in across the embedding networks.
Intriguingly enough, the result of a measurement at one position  dictates the possible outcome of a measurement performed at a different location.  Stochastic trajectories appear reciprocally {\it entangled} over arbitrary distances, a subtle effect instigated by endogenous finite size corrections which is completely lost by mean-field analysis. This is the first time that such long-ranged stochastic patterns are observed
in a simple reaction model, and we conjecture that they could represent a new route for information processing in complex networks. As we shall argue in the following, the interplay between finite-size effects and non-linear coupling sits at the root of the observed phenomena.  

As a first step, we introduce the minimal model for finite excitatory and inhibitory populations on a single mesoscopic node (a ``patch''). We next turn 
to consider the coupled dynamics of a multi-species, excitatory and inhibitory model distributed on a network. 
The simple setting where just two nodes are considered will serve as a basis to develop the main tools for our analysis.

{\bf Excitatory and inhibitory dynamics on an isolated patch.}   
In our single-node model, we label excitatory and inhibitory elements by, respectively $X$ and $Y$. They live in a mean-field interacting
patch (node) of volume $V$, and undergo the following reactions:

\begin{equation}
\begin{array}{lcl}
\emptyset & \overset{f[s_x]}{\longrightarrow} & X \\
X & \overset{1}{\longrightarrow} & \emptyset\\
\emptyset & \overset{f[s_y]}{\longrightarrow} & Y \\
Y & \overset{1}{\longrightarrow} & \emptyset 
\end{array}
\label{chemical}
\end{equation}

where $\emptyset$ denotes an infinite reservoir, $f(s)\!=\!1/(1\!+\!\exp (\!-\!s))$ is a sigmoid function and 
$s_x  \!=\!   \!-\! r \left( \frac{n_Y}{V} \!-\! \frac{1}{2} \right)$,  
$s_y  \!=\!    r \left( \frac{n_X}{V} \!-\! \frac{1}{2} \right)$. Here,  $r\!>\!0$ is the only free parameter; $n_X$ and $n_Y$ respectively identify the number of elements of type $X$ and $Y$. From the above definition of $s_{x,y}$, and $f(\cdot)$, it is immediately clear that a large population of inhibitory elements suppresses the excitatory population; conversely, a large excitatory population enhances the inhibitory one.

\begin{figure}
 \centering
   {\includegraphics[width=7cm]{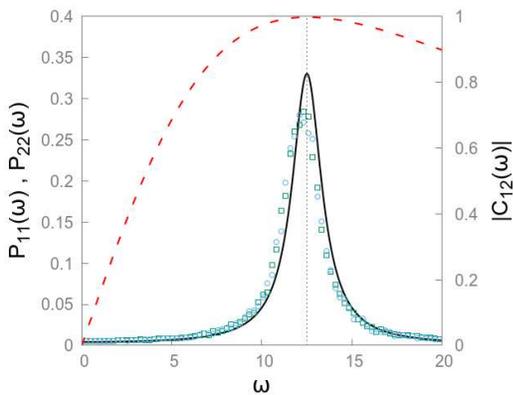}}
   \caption{The theoretical power spectrum $P_{11}(\omega)=P_{22}(\omega)$ (\ref{theo}) is depicted with a solid line. Symbols refer to the power spectra computed from averaging independent realization 
   of the Gillespie dynamics. The squares refers to excitators ($X$), while the circles stand for inhibitors ($Y$). The vertical dotted line is traced at $\omega=\lambda_{Im}=\sqrt{r}/4$. The theoretical dashed line represents $|C_{12}|=|C_{21}|$. A phase lag equal to $\pi/2$ is predicted. Here $r=50$ so to allow for the isolated peak in the power spectra to be distinctively revealed. }
   \label{fig1}
  \end{figure}

Introduce $P_{\boldsymbol n}(t)$ to label the the probability for the system to be in state ${\boldsymbol n } = (n_X,n_Y)$ at time $t$. Transitions from one state to another are caused by the above chemical equations. $T({\boldsymbol n} | {\boldsymbol n}')$ is the transition rate from state 
${\boldsymbol n} $ to state ${\boldsymbol n} '$, compatible with the former. The dynamics of the system is governed by a master equation which takes the generic form
$\frac{\mathrm{d}}{\mathrm{d}t} P_{\boldsymbol n}(t) \!=\! \sum_{\boldsymbol n'} T \left( {\boldsymbol n} | {\boldsymbol n'} \right) P _{\boldsymbol n'}(t) \!-\! T \left( {\boldsymbol n} '| {\boldsymbol n }\right) P _{\boldsymbol n}(t)$.
To progress in the analytical understanding of the problem, one can write down the equation for the average quantities $\langle n_X \rangle \!=\! \sum_{\boldsymbol n} n_X P _{\boldsymbol n}$ and $\langle n_Y \rangle \!= \! \sum_{\boldsymbol n} n_Y P _{\boldsymbol n}$.  In the limit where the volume $V$ (hence the number of constituents) is large, endogenous fluctuations fade away:  the system can be described in terms of the concentrations of the chemical species $x \! =\! \lim_{V \rightarrow \infty} \langle n_X \rangle/V$ and  $y \!=\!\lim_{V \rightarrow \infty} \langle n_Y \rangle/V$. In doing so one eventually gets:
\begin{eqnarray}
\dot{x} &=& - x + f \left( - r ( y - \frac{1}{2} ) \right) \\
\dot{y} &=& - y + f \left( r ( x - \frac{1}{2} ) \right) \nonumber
\label{deterministic_eq}
\end{eqnarray}
The above system admits a single nontrivial fixed point (or steady state) 
$x_f \!=\! y_f \!=\! \frac{1}{2}$ \cite{footnote1}. It is straightforward to characterize the stability of $(x_f,y_f)$ by computing the eigenvalues $\lambda$ of the Jacobian matrix $\mathbf{J}$ associated to system (\ref{deterministic_eq}), evaluated at equilibrium. Performing the calculation one gets $\lambda \!=\! \lambda_{Re} \!+\! i \lambda_{Im} \!=\! -\! 1 \! \pm \! i \sqrt{r}/4$. The real part of $\lambda$ is negative and the fixed point is therefore stable. Furthermore, the eigenvalues are complex: 
stochastic oscillations sustained by the endogenous noise  can eventually set in around the fixed point. Under the standard linear noise approximation (LNA)~\cite{vankampen}, finite-size endogenous stochastic effects act as linear deviations from the deterministic solution. More specifically, one stipulates $x(t) \!=\! x_f \!+\! V^{-1/2} \xi_1$ and $y(t) \! = \! y_f \!+\! V^{-1/2} \xi_2$, where 
$\mathbf{\xi} \! = \! (\xi_1,\xi_2)$ stands for the stochastic perturbation.  The LNA assumes that $V$ to be large, so that only linear terms in $\mathbf{\xi}$ are to be retained when the above ansatz is inserted in the governing master equation. The factor  $V^{-1/2}$ reflects the Gaussian nature of the approximation. The linearized fluctuations can be shown to obey a Langevin equations~\cite{biancalani} in the form 
$\dot{\xi_{i}} \!=\! \sum_{j} J_{ij} {\xi} \!+\! \eta_i$, where $\eta_i(t)$ is a Gaussian noise term with zero mean and with correlator $<\eta_i(t)\eta_j(t')> \! =\! \delta_{ij} \delta (t-t')$. Denoting the Fourier transform of the $\xi_i(t)$ as
$\tilde{\xi_i}(\omega)$, one readily gets $\tilde{\xi}_i (\omega) \!=\! \sum^2_{j=1} \Phi^{-1}_{ij}(\omega) \tilde{\eta}_j(\omega)$ where  $\Phi_{ij} \!= \! - \! J_{ij} \! - \! i \omega \delta_{ij}$. Endogenous oscillations can be analyzed by computing the power spectral density matrix (PSDM):
\begin{equation}
\label{theo}
P_{ij} (\omega)= <\tilde{\xi}_j(\omega)\tilde{\xi}_j^*(\omega)> = \sum^2_{l=1}\sum^2_{m=1} \Phi^{-1}_{il} (\omega)\delta_{lm} \left( \Phi^{\dag} \right)^{-1}_{mj}(\omega)
\end{equation}
The diagonal entries of the PSDM are real and coincide with the power spectra for the fluctuations, associated to each species. The (generally complex) off-diagonal elements of the PSDM can be properly normalized so to yield the Complex Coherence Function (CCF) $C_{ij} ( \omega ) \!=\! \frac{P_{ij}(\omega)}{\sqrt{P_{ii}(\omega)P_{jj}(\omega)}}$. 
As explained in~\cite{rozhnova,challenger} the magnitude $|C_{ij}|$ measures the degree of correlation between two signals, as a function of $\omega$. The phase $\phi_{ij}\!=\!\arctan[(C_{ij})_{Im}/(C_{ij})_{Re}]$ quantifies the phase lag between the two inspected signals. In Figure  \ref{fig1} we depict the power spectra $P_{11}$ and  $P_{22}$ (scale on the left), together with  $|C_{12}|\!=\!|C_{21}|$ (scale on the right). The power spectra display an identical profile (due to the symmetry of the equations) which is peaked at $\omega \! \simeq \! \lambda_{Im}$: the  endogenous fluctuations gets amplified through a resonant mechanism that yields quasi-cycle oscillations.  Symbols in the upper panel of Figure \ref{fig1} refer to the numerically computed power spectra and confirm the adequacy of the linear noise calculation.  The magnitude of  $|C_{12}|$ is maximum, when the power spectra are.  At this point the phase lag between the oscillators,  the excitators and the inhibitors, abuts on $\pi/2$. The theory predictions are fully confirmed by direct Gillespie \cite{gillespie} based simulations (see SI) of reactions (\ref{chemical}).
Starting from this setting, and building on the methodology that we have here briefly revised, we shall proceed to study the issue of synchronization when two or more replica of model (\ref{chemical}) are coupled together. 
  
{\bf A many patches network model of coupled excitatory-inhibitory dynamics.}  
We now turn to considering an immediate generalization of the above model to the case where excitators and inhibitors populations are bound to occupy different spatial positions and
proceed to study the issue of synchronization when two or more replica of model (\ref{chemical}) are coupled together by a complex network topology. 
Symbols used to identify individual entities are now decorated with an additional index, so to specify the node to which they refer to.  More concretely, we now deal with the elements $X_k$ and $Y_k$, with $k\!=\!1,2,\ldots,\Omega$. Each element is subject to birth and death chemical equations of the type illustrated above~\cite{rozhnova,challenger}, see equations \ref{chemical}, but different connected nodes interact diffusively via their concentrations gradient, which modifies the inputs $s_{x_i}$, $s_{y_i}$ of the non linear rate function $f(\cdot)$. More concretely, in the simple case of only two nodes, 
with an obvious meaning of the notation involved we have
$s_{x_1}  =   - r \left( \frac{n_{Y_1}}{V} - \frac{1}{2} \right)+D \Delta_{1\,2}$,
$s_{y_1}  =     r \left( \frac{n_{X_1}}{V} - \frac{1}{2} \right)+ D \Delta_{1\,2}$,
$s_{x_2}  =   - r \left( \frac{n_{Y_2}}{V} - \frac{1}{2} \right)-  D \Delta_{1\,2}$,
$s_{y_2}  =     r \left( \frac{n_{X_2}}{V} - \frac{1}{2} \right)-  D \Delta_{1\,2}$
where $\Delta_{1\,2}=(\frac{n_{X_2}}{V}-\frac{n_{X_1}}{V})-(\frac{n_{Y_2}}{V}-\frac{n_{Y_1}}{V})$ and $D$ is our coupling parameter. 
This setting could be easily generalized to account for a network of $\Omega$ nodes by introducing the associated adjacency matrix $\mathbf{A}$: $A_{ij}\!=\!1$ if nodes $i$ and $j$ are connected, $A_{ij}\!=\!0$ otherwise.  
In the following we will consider symmetric coupling $A_{ij}\!=\!A_{ji}$, but we anticipate that our conclusions remain unchanged when asymmetric couplings are allowed for. Generalizing the expression introduced above for the case of two nodes, we set
$s_{x_i} = -  r \left( \frac{n_{Y_i}}{V} - \frac{1}{2} \right) + D \sum_j^{\Omega}\Gamma_{ij} ( \frac{n_{X_j}}{V}   \!-\! \frac{n_{Y_j}}{V})$ and  
$s_{y_i} =  r \left( \frac{n_{X_i}}{V}  - \frac{1}{2} \right) + D \sum_j^{\Omega}\Gamma_{ij} (\frac{n_{X_j}}{V}   \!-\! \frac{n_{Y_j}}{V}))$ 
 where $\Gamma_{ij} \!=\! A_{ij} \!-\! \kappa_i \delta_{ij}$ is the standard discrete Laplacian operator and $\kappa_i$ stands for the connectivity of node $i$. 

The state of the system is photographed by the $2 \Omega$ components concentrations vector ${\boldsymbol n } \!=\! (n_{X_1},n_{Y_1},n_{X_2},n_{Y_2}, \ldots, n_{X_\Omega},n_{Y_\Omega})$. The dynamics of the stochastic system  is still ruled by a master equation, under the Markov assumption. In the limit $V \rightarrow \infty$, one can introduce the mean field concentrations $(x_1,y_1,x_2,y_2, \ldots, x_\Omega,y_\Omega)$. The ODEs 
that govern the time evolution of the deterministic variables constitute the natural generalization of equations (\ref{deterministic_eq}). Remark that the imposed coupling senses the gradient of concentrations, for both excitatory and inhibitory populations. It is therefore immediate to realize that $x_i\!=\!x_f$ and $y_i\!=\!y_f$ $\forall i$ is the fixed point of the deterministic model. To illustrate our approach, we first concentrate on a two nodes network. 

The stability of the fixed point can be assessed, by computing the eigenvalues of the $4 \times 4$ Jacobian matrix $\mathbf{J}$. Two eigenvalues coincide with the ones calculated above for the isolated patch setting, $\lambda_{1,2}\!=\!-\!1 \pm i (r/4)$. The two additional eigenvalues read  $\lambda_{3,4}\!=\!\pm \sqrt{\frac{r}{4}(D\!-\!\frac{r}{4})} \!-\! 1$ and depend on the coupling strength $D$ \cite{footnote2}. For  $D\!>\!D_c\!=\!r/4 \!+\! 4/r$, $\lambda_{3}$ is real and positive and the fixed point turns therefore unstable in a pitchfork bifurcation (see SI). For $D\!<\!D_c$, however the solution $x_1\!=\!x_2\!=\!x_f\!=\!1/2$ and $y_1\!=\!y_2\!=\!y_f\!=\!1/2$ is stable and the thermodynamic limit system displays a uniform level of activity, for both excitators and inhibitors, across the two nodes. 
When finite size effects are considered, excitators (reps. inhibitors) populations on each node executes quasi-regular oscillations about the trivial deterministic fixed point. Nodes are formally decoupled when $D$ is set to zero, so that stochastic trajectories on distinct nodes are disentangled. 
Conversely, when $0 \!<\! D \!<\!D_c$  inter-nodes species are effectively coupled, the degree of reciprocal influence being more pronounced the closer $D$ is to $D_c$. 
\begin{figure}[t!]
 \centering
\includegraphics[width=8.5cm]{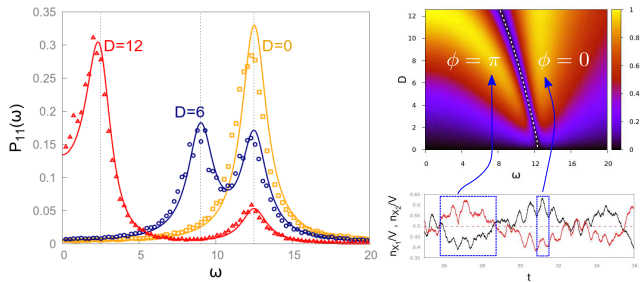}
   \caption{Left panel: the power spectrum $P_{11}$ is plotted as function of $\omega$, for different choices of $D$. Lines refer to the theoretical predictions. Symbols are obtained by averaging over many realizations of the stochastic simulations. Right panel: $|C_{13}|$ is plotted in the plane $(\omega, D)$. Two regions can be identified where the synchronization take place. These are separated by the dashed (white) line, obtained by setting $|C_{13}|\!=\!0$. The synchronization at small frequencies occur in anti-phase ($\phi\!=\!\pi$), while at high frequencies the theory predicts  $\phi\!=\!0$. The stochastic trajectories ($n_{X_1}$ and $n_{X_3}$ vs. time $t$) confirm the adequacy of the LNA. In phase and anti-phase regimes of synchronization are highlighted in the boxes. Here, $r\!=\!50$ and $V\!=\!20000$.}
   \label{fig2}
  \end{figure}
Can the imposed coupling enforce a synchronization of the emergent stochastic oscillations? This is the question that we are going to answer hereafter, building on the methodology illustrated above and computing the $4 \times 4$ PSDM  associated to the system at hand. Focus first on the diagonal elements of the PSDM , i.e. the power spectra $P_{ii}(\omega)$, $i\!=\!1,\dots,4$. The result of the analysis are plotted in the left panel of Figure (\ref{fig2}), for $i\!=\!1$. When $D\!=\!0$, the power spectrum displays an isolated peak, located at $\omega \simeq r/4$ (rightmost vertical dashed line) in agreement with the analysis carried out for the single patch case study. When $D$ increases a second peak develops, and progressively gains in magnitude. Its position is adequately captured by the (positive) imaginary component of the eigenvalues $\lambda_{3,4}$ (dashed lines). When $D$ approaches the critical threshold $D_c$, the leftmost peak stands alone, and the other fades away. For intermediate parameter settings, the stochastic oscillators are forged by the simultaneous presence of two leading frequency, whose relative importance can be controlled as wished. Gillespie bases simulations of the coupled reactions, performed for different values of $D$, confirm the correctness of the theory.  Similar observations apply to $i\!=\!2,3,4$. 

Consider now the off diagonal entries of the PSDM and build the corresponding CCF.  To shed light onto the inter-nodes correlation between excitators, we plot in the right panel of Figure (\ref{fig2}) $|C_{13}|$, in the plane $(\omega, D)$, using an apt color code. For small $D$, the signals are, as expected, completely independent. By increasing $D$, two regions are found where $|C_{13}|$ takes values close to unit.  Quasi-cycles displayed by the excitatory populations attached to distinct nodes do synchronize, for sufficiently large values of the coupling strength. Intriguingly enough, and at odds with the examples so far reported in the literature, the synchronization is established for two different characteristic frequencies. These are the indirect reflex of the two peaks identified in the power spectrum. Even more importantly, the two aforementioned regions are separated by a distinct frontier (white dashed line) where $|C_{13}|$ is found to be identically equal to zero: in the left portion of the plan, with respect to the white dashed separatrix, the phase lag is {\it exactly} $\pi$. The stochastic trajectories are hence predicted to be in anti-phase, on short frequencies, or equivalently, long periods. In the complementary portion of the plane, i.e. on the right of the separatrix, $\phi$ is found to be zero, thus implying perfect synchronization at large frequencies or short periods. Direct simulations confirm a posteriori the scenario depicted above, see trajectories annexed to the right panel of Figure (\ref{fig2}).
Stochastic trajectories referred to the same species attached to contiguous nodes are {\it entangled}. The result of a measurement at one node roughly determines the outcome of a measurement simultaneously performed at the other node. Leaving aside more fundamental reflections,  we remark that such entangled states can be hierarchically assembled to yield macroscopic patterns, as we shall demonstrate herafter. 
Complementary information can be drawn by inspecting the other off-diagonal elements of the PSDM, see SI.

We are now in a position to extend the analysis to the relevant case where the interaction between excitators and inhibitors is mediated by a complex network. As outlined for the two nodes setting, the interactions is supposed to be diffusive in that it senses the difference of concentrations between homologous species on distinct nodes. Furthermore, the coupling is embedded in the nonlinear function $f(\cdot)$, here introduced to exemplify the activation process.  

We first determine the stability of the uniform fixed point $x_i\!=\!x_f$ and $y_i\!=\!y_f$ $\forall i$.
To do so, one can introduce a small non homogenous perturbation, and solve the linearized equation by expanding the perturbation on the basis of the eigenvectors of the Laplacian $\Gamma$. The analysis is carried out in the SI, and returns a closed analytical expression for the critical coupling, namely $D_c\!=\! (16/r \!+\! r)/(2 \max_{\alpha} \left|  \Lambda^{(\alpha)}\right|)$. Here, $\Lambda^{(\alpha)}$,  $\alpha\!=\!1, \dots, \Omega$, denote the eigenvalues of the Laplacian operator. Remark that this latter formula reduces to the one obtained above when $\Omega\!=\!2$.  

As a first example we consider a linear ring made of $\Omega\!=\!4$ nodes and calculate the $2 \Omega \! \times\! 2 \Omega$ elements of the PSDM.  A sequential alternation of phase and anti-phase synchronization is predicted for the stochastic excitatory signals, registered across the ring (see SI). Similar conclusions can be drawn when probing the degree of synchronization between excitators and inhibitors on different nodes.
Patterns of activation instigated by the noise assisted drive towards self-organization are hence expected to emerge. In the left panel of Figure \ref{fig3} a snapshot of the stochastic dynamics is displayed which supports this conclusion:  a regular sequence of active/inactive excitators/inhibitors is found, when circulating along the chain. Excitators (upper circle, continuous arrow) and inhibitors (lower circle, dashed arrow)  are in anti-phase on the node they happen to share:  arrows point upward (resp. downward) if the measured activity is more (resp. less) pronounced as compared to the mean field uniform equilibrium. The system keeps on switching between the configuration depicted in  Figure \ref{fig3}, and its negative analogue, where the inactive populations on a given node turn active, and viceversa (see full movie, annexed as supplementary material).  The architecture of the network plays indeed a crucial role. Robust patterns which exploit the phase/anti-phase dichotomy on a closed ring, necessarily require accommodating for an even number of nodes. When nodes are odd, frustration may occur \cite{gollo}. This is an intriguing effect on which we shall report elsewhere. Recall that the formalism here developed applies to generic networks, not just to regular lattices. For demonstrative purposes, we show in the right panel of Figure \ref{fig3} a snapshot of the  (excitatory) activation pattern obtained when the system is placed on a tree with branching ratio equal to $4$ (see SI for a movie of the dynamics).  

Summing up, we have here proved that endogenous
noise promotes a coordinated pattern of action, in a minimalistic
model of excitatory and inhibitory interactions -- showing no Hopf bifurcations in the thermodynamic, deterministic limit.
This latter model encompasses, for its inherent simplicity, a
large gallery life science applications, ranging from neuroscience
to the study of genetic circuits. Noise induces quasi-cyclic oscillations are long-ranged correlated in anti-phase stochastic patterns,
 inter-tangled motifs that could indeed convey important tips on how living systems handle computational tasks and information processing.

\begin{figure}[t]
 \centering
\includegraphics[width=9cm]{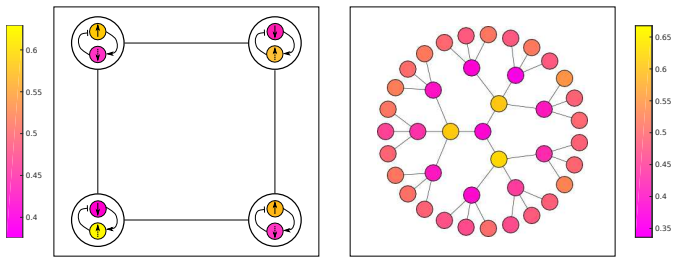}
   \caption{Snapshot of the stochastic dynamics for two networks topologies: the stochastic density of the excitators is plotted. Left panel: a chain of four nodes is considered. In each node the density of excitators (upper circles) and inhibitors (lower circles)  is depicted. Right panel: a snapshot of the stochastic pattern of activation of the excitators is shown.}
   \label{fig3}
  \end{figure}
\vspace{-0.3 cm}
\section*{Acknowledgments}
We have benefited from discussions with A. Politi. The authors acknowledge financial support from H2020-MSCA-ITN-2015 project COSMOS  642563. 

\appendix

\section{Excitatory and inhibitory dynamics on an isolated patch: phase shift of $\pi/2$.}   

\begin{figure}[!h]
 \centering
   {\includegraphics[width=8cm]{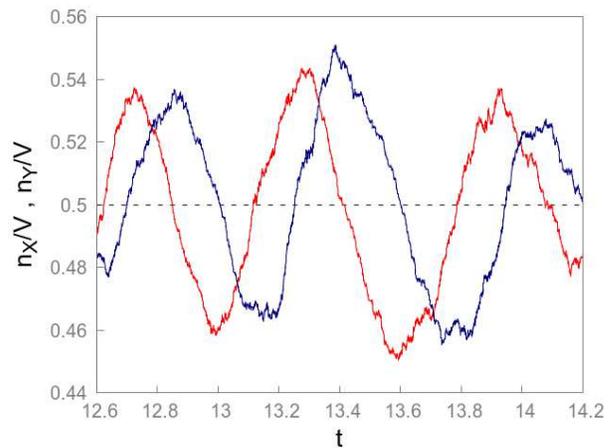}}
   \caption{Stochastic trajectories ($n_X$ and $n_Y$ versus time $t$).  The fixed point of the underlying deterministic model is depicted with a dashed black line.}
   \label{suppl_fig1}
  \end{figure}

Recall from the main text that, for a single patch model,  the magnitude of  $|C_{12}|$ is maximum, when the power spectra are.   The phase lag between the oscillations displayed by excitators and inhibitors is predicted  $\pi/2$. To check the validity of the theory we performed a direct integration of system (\ref{chemical}) in the main text, by means of the celebrated Gillespie algorithm. Results of the analysis are reported in Figure  \ref{suppl_fig1}: the concentrations of $n_X/V$ and $n_Y/V$ are indeed synchronized with a phase shift of about $\frac{\pi}{2}$, in agreement with the theoretical predictions.

\section{The two patches model}

In this section we provide supplementary information to complement the description of the two patches model, as introduced in the main paper. Individual elements of the excitatory and inhibitory species are respectively labeled $X_i$
and $Y_i$. The index $i$ identifies the nodes to which the elements refer to. The stochastic dynamics of the system can be formulated in terms of chemical equations:

\begin{equation}
\begin{array}{lcl}
\emptyset & \overset{f[s_x^i]}{\longrightarrow} & X_i \quad i = 1, 2\\
X_i & \overset{1}{\longrightarrow} & \emptyset\\
\emptyset & \overset{f[s_y^i]}{\longrightarrow} & Y_i \quad i = 1, 2 \\
Y_i & \overset{1}{\longrightarrow} & \emptyset 
\end{array}
\label{chemical}
\end{equation}
where $s_x^i$ and $s_y^i$ read: 
\begin{equation}
\begin{array}{lcl}
s_x^i & =  & - r \left( \frac{n_{Y_i}}{V} - \frac{1}{2} \right) + D \left( \frac{n_{X_j}}{V} - \frac{n_{X_i}}{V} \right) - D \left( \frac{n_{Y_j}}{V} - \frac{n_{Y_i}}{V} \right) \\
s_y^i & =  & r \left( \frac{n_{X_i}}{V} - \frac{1}{2} \right) + D \left( \frac{n_{X_j}}{V} - \frac{n_{X_i}}{V} \right) - D \left( \frac{n_{Y_j}}{V} - \frac{n_{Y_i}}{V} \right)
\end{array}
\label{currents}
\end{equation}

Here $D$ stands for the coupling parameter. As stated in the main body of the paper, the imposed coupling senses the gradient of concentrations, for both excitatory and inhibitory populations.
The state of the system can be tracked via $\mathbf{n} = ( n_{X_1}, n_{Y_1}, n_{X_2}, n_{Y_2})$, namely by quantifying the number of discrete elements belonging to each species, on every node.  The evolution of the probability 
$P_{\boldsymbol n}(t)$ of seeing the system in state ${\boldsymbol n}$ at time $t$, is governed by a master equation which takes the general form:
\begin{equation}
\frac{\mathrm{d}}{\mathrm{d}t} P_{\boldsymbol n}(t) = \sum_{\boldsymbol n'} T \left( {\boldsymbol n} | {\boldsymbol n'} \right) P _{\boldsymbol n'}(t) - T \left( {\boldsymbol n} '| {\boldsymbol n }\right) P _{\boldsymbol n}(t).
\label{ME}
\end{equation}

Starting from this setting, one can recover the equations that rule the deterministic, mean field dynamics and moreover elaborate on the role played by the endogenous fluctuations.

\subsection{Deterministic limit}

\par In the limit where the volume $V$  is large, endogenous fluctuations vanish.  The system can be described in terms of the concentrations of the chemical species 
$x_i =\lim_{V \rightarrow \infty} \langle n_{X_i} \rangle/V$ and  $y_i =\lim_{V \rightarrow \infty} \langle n_{Y_i} \rangle/V$. It is then straightforward to recover the following ODEs for the coupled evolution of the mean field concentrations:

\begin{equation}
\begin{array}{lcl}
\dot{x_1} & = & -  x_1 + f \left( - r ( y_1 - \frac{1}{2} ) + D (x_2 - x_1 ) - D (y_2-y_1)\right) \\
\dot{y_1} & = &  - y_1 + f \left( r ( x_1 - \frac{1}{2} ) + D (x_2 - x_1 ) - D (y_2-y_1) \right) \\
\dot{x_2} & = & -  x_2 + f \left( - r ( y_2 - \frac{1}{2} )  + D (x_1 - x_2 ) - D (y_1-y_2)\right) \\
\dot{y_2} & = &  - y_2 + f \left( r ( x_2 - \frac{1}{2}  ) + D (x_1 - x_2 ) - D (y_1-y_2) \right) 
\end{array}
\label{mean_field2}
\end{equation}

\begin{figure}[!h]
 \centering
 \includegraphics[width=8cm]{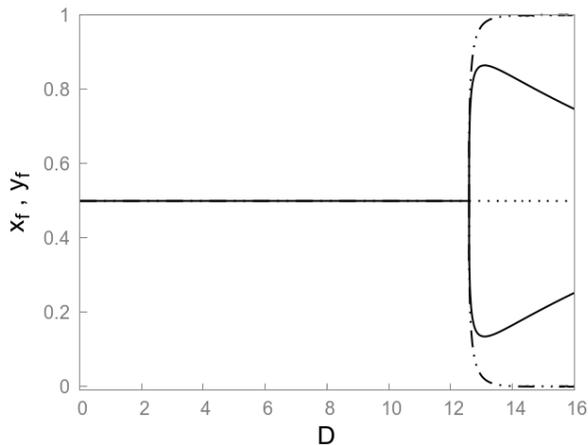}
   \caption{Bifurcation diagram for the two nodes system, in its deterministic version. Above the bifurcation point ($D>D_c$), the solid black lines refer to the excitators $x_{1,2}$, while the dashed-dotted lines stand for the inhibitors $y_{1,2}$.}
   \label{suppl_fig2}
  \end{figure}

The above system admits a uniform fixed point  $x_i=x_f =1/2$ and  $y_i=y_f =1/2$. This latter fixed point is stable for  $D\!<\!D_c\!=\!r/4 \!+\! 4/r$. At $D=D_c$, the system undergoes a pitchfork bifurcation: two symmetric stable branches appear, as illustrated in Figure   \ref{suppl_fig2}.

To gain further insight into the dynamics of the system, we perform a linear expansion close to the bifurcation point, $D=D_c=r/4+4/r$.  More specifically, set $x_1 = x_f - \epsilon_x$, $x_2 = x_f + \epsilon_x$, $y_1 = y_f - \epsilon_y$, $y_2 = y_f + \epsilon_y$. By inserting the above ansatz into the full non linear equations (\ref{mean_field2}) and performing a linear expansion in the perturbation parameters ($\epsilon_x,\epsilon_y$) eventually yields:
 
\begin{equation}
\epsilon_x = - \epsilon_y \frac{\frac{r}{8} - \frac{2}{r}}{1 + \frac{D_c}{2}}  \nonumber
\end{equation}

a closed expression that allow to immediately appreciate the relative modulation of the fixed points above the instability threshold.
In particular for $r>4$, $\epsilon_x$ and $\epsilon_x$ have opposite signs.  Imagine that the concentration of the excitators on node $1$ displays a level of activity that is larger than $1/2$. In other words $x_1$ belongs to the upper stable (solid) branch in Figure \ref{suppl_fig2}. Then, $x_2$ is forcefully associated to the lower branch of the bifurcation diagram;  $y_1$ and $y_2$ display in turn an opposite internal arrangement. In other words, above the bifurcation point, the deterministic system manifests a degree of  spatial organization (across nodes) that, to some extent, recalls the noise driven  motifs found for $D<D_c$, as  discussed in the main body of the paper. Remark however that deterministic patterns are stationary, at variance with stochastic ones. Endogenous noise forces in fact the stochastic system to continuously blink between different states, respecting local and long-ranged intertangled constraints.

\subsection{Linear noise approximation and PSDM}

To elaborate on the role played by finite size (endogenous) fluctuations we operate under the linear noise approximation.  In complete analogy with the case of an isolated patch we introduce the following ansatz $x_1(t) = x_f +\xi_1/\sqrt{V}$, $y_1(t) = y_f + \xi_2/\sqrt{V}$, $x_2(t) = x_f + \xi_3/\sqrt{V}$, $y_2(t) = y_f + \xi_4/\sqrt{V}$, where  $\xi_i$ with $i=1, \cdots 4$ stand for the stochastic corrections to the mean field equilibrium.
We then insert the above ansatz into the master equation and expand in series of $1/\sqrt{V}$. At the leading order of approximation we recover the mean field equations (\ref{mean_field2}). At the next to leading order,  
one ends up with the following linear Langevin equations ($i=1,\dots,4$):
\begin{equation}
\frac{\mathrm{d}}{\mathrm{d}t} \mathbb{\xi}_i = \sum^{4}_{j=1} J_{ji} \xi_i + \eta_i
\end{equation}
where $\mathbf{J}$ stands for the Jacobian $4\times 4$ matrix associated to the deterministic system (\ref{mean_field2}).  Here, $\eta_i$ is Gaussian white noise with $< \eta_i (t) \eta_j(t') >  = \delta_{ij} \delta (t-t')$. Denote by $\tilde{\xi}_i(\omega)$ the Fourier transform of $\xi_i(t)$. Then, we get
\begin{equation}
\tilde{\xi}_i(\omega) = \sum^4_{j=1} \Phi_{ij}^{-1}(\omega)\tilde{\eta}_j(\omega)
\end{equation}
where $\Phi_{ij} = -J_{ij}-\omega \delta_{ij}$. To determine the  PSDM  we compute  the elements $P_{ij}(\omega) = <\tilde{\xi}_i(\omega)\tilde{\xi^*}_j(\omega)>$. Since the two nodes share identical parameters, the PSDM 
can be completely characterized in terms of  $6$ different entries, namely $P_{11}(\omega) = P_{33}(\omega)$, $P_{22}(\omega) = P_{44}(\omega)$, $P_{12}(\omega) = P_{34}(\omega) $, $P_{13}(\omega)$, $ P_{24}(\omega)$ , $P_{14}(\omega) = P_{23}(\omega)$. Following the discussion reported in the main text, we introduce the rescaled parameters $C_{ij}(\omega) = \frac{P_{ij}(\omega)}{\sqrt{P_{ii}(\omega)P_{jj}(\omega)}}$. The magnitude of $C_{13}$ is plotted in the right panel of Figure 2 in the main body of the paper. In the following, we display the other relevant coefficients and briefly elaborate on their intrinsic meaning. 

In the left panel of Figure  \ref{fig3} we report the magnitude of $C_{12}$ in the reference plane $(D,\omega)$. When $D = 0$, excitators and
inhibitors belonging to the same node oscillate with a phase lag that we already have quantified $\pi/2$. This
condition is perpetrated, when $D$ is made to increase inside the region of interest ($D < Dc$). Contextually,
a second branch emerges, where $|C_{12}|$ takes significant
values. The phase lag gets progressively modulated along
the newly produced branch and eventually approaches $\pi$, for large enough coupling amount. Also in this case,
numerical experiments are found in excellent agreement
with the theory predictions. Similar conclusions
can be drawn by analyzing $C_{24}$ and $C_{34}$. In the right panel of Figure  \ref{fig3} the magnitude of $|C_{14}|$ in plotted in the plane $(D,\omega)$.

\begin{figure}[!h]
 \centering
   {\includegraphics[width=8cm]{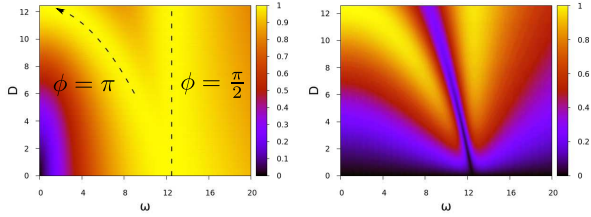}}
   \caption{Left panel: map describing the magnitude of $C_{12}$ in the reference plane $(D,\omega)$. This parameter allows to resolve the degree of synchronization between excitators and inhibitors in the same node. Right panel:
   $|C_{14}|$ in plotted in the plane $(D,\omega)$.}
   \label{fig3}
  \end{figure}

\section{Excitatory and inhibitory dynamics on a complex network.}
\par We begin this section by determining a closed analytical expression for the critical value of the coupling constant $D$ that sets the limit of stability of the homogeneous fixed point. To this end we consider 
a network made of $\Omega$ nodes. The architecture of the network is specified by the adjacency matrix  $\mathbf{A}$. The state of the system in its deterministic limit is in turn described by a $2 \Omega$ vector $ \mathbf{z}=(x_1,y_1,x_2,y_2,\dots,x_{\Omega},y_{\Omega})$. The deterministic equations read: 
\begin{eqnarray*}
\dot{x_i} &=& -x_i + f \left( - r\left( y_i - \frac{1}{2} \right) + D \sum_j^{\Omega}\Gamma_{ij}x_j  - D \sum_j^{\Omega}\Gamma_{ij}y_j \right) \\
\dot{y_i} &=& -y_i + f \left(r\left( x_i - \frac{1}{2} \right) + D \sum_j^{\Omega}\Gamma_{ij}x_j  - D \sum_j^{\Omega}\Gamma_{ij}y_j \right) 
\end{eqnarray*}
where $\Gamma_{ij} = A_{ij} - \kappa_i \delta_{ij}$ are the entries if the discrete Laplacian operator, and $\kappa_i$ is the connectivity of node $i$
Set $x_i = x_f + \delta x_i$ and $y_i = y_f + \delta y_i$, insert these expressions in the deterministic equations and expand at the linear order in the imposed perturbation $(\delta x_i,\delta y_i)$: 

\begin{eqnarray*}
\dot{\delta x_i} &=& - \delta x_i + \left( - r \delta y_i + D \sum^{\Omega}_j \Gamma_{ij} \delta x_j - D \sum^{\Omega}_j \Gamma_{ij} \delta y_j \right) f'(\mathbf{z}) \\ \nonumber
\dot{\delta y_i} &=& -\delta y_i + \left( r \delta x_i + D \sum^{\Omega}_j \Gamma_{ij} \delta x_j - D \sum^{\Omega}_j \Gamma_{ij} \delta y_j \right) f'(\mathbf{z})
\label{pert}
\end{eqnarray*}

Where $f'(\cdot)$ denotes the derivative of the sigmoid function $f(\cdot)$. It is immediate to realize that $f'(\mathbf{z}) = f(\mathbf{z}) (1-f(\mathbf{z}))=\frac{1}{4}$. We then set to
expand the perturbations $(\delta x_i$, $\delta y_i)$ on the basis of the eigenvectors of the Laplacian $\Gamma$. To this end, denote by $\Lambda^{(\alpha)}$ (with $\alpha = 1 \dots \Omega$) the eigenvalues of $\Gamma$ $(\alpha = 1 \dots \Omega$) and by ${\Phi_i^{(\alpha)}}$ the associated eigenvectors, namely $\sum_j \Gamma_{ij}\Phi_j^{(\alpha)} = \Lambda^{(\alpha)} \Phi_i^{(\alpha)}$.
In formulae, we require:  
\begin{subequations}
\begin{equation}
\delta x_i = \sum^{\Omega}_{\alpha} c_{\alpha} \exp (\lambda_{\alpha}) \Phi_i^{(\alpha)}
\end{equation}
\begin{equation}
\delta y_i = \sum^{\Omega}_{\alpha} b_{\alpha} \exp (\lambda_{\alpha}) \Phi_i^{(\alpha)}
\end{equation}
\end{subequations}
Inserting in the equations for the perturbation,  carrying out the calculation and projecting on each independent eigendirection, one eventually ends up with.
\[
\begin{array}{cccc}
\left[
\begin{array}{cc}
-1 + \frac{D \Lambda^{(\alpha)}}{4} - \lambda_{\alpha} & -\frac{1}{4}(r +D \Lambda^{(\alpha)}) \\
  \frac{1}{4}(r+D \Lambda^{(\alpha)}) & -1 -  \frac{ D \Lambda^{(\alpha)}}{4} - \lambda_{\alpha} 
\end{array}\right]
& 
\left[\begin{array}{c}
c_{\alpha} \\
b_{\alpha}
\end{array}\right]
&
=
&
0
\end{array}
\]
The above homogeneous system admits a non trivial solution provided the matrix in square brackets has zero determinant. This latter condition yields a second order equation for $\lambda_{\alpha}$ as function of $\Lambda^{(\alpha)}$ and $D$. 
Consider then  $\lambda_{\alpha}^{+}$, the largest of the two roots. If the real part of $\lambda_{\alpha}^{+}$ is positive, then the perturbation grows exponentially and the homogeneous fixed point is unstable. In our analysis we considered symmetric networks: in this case the eigenvalues $\Lambda_{\alpha}$ are real and semi-negative defined. 
After a  straightforward manipulation it is immediate to conclude that $(\lambda_{\alpha}^{+})_{Re}<0$ provided $D$ is smaller than the critical value:

\begin{equation}
D_c = \frac{\frac{16}{w} + w}{2 \max\limits_{\alpha}|\Lambda^{(\alpha)}|} 
\end{equation}

This latter expression has been successfully validated against numerical inspection. For $D<D_c$ the homogeneous fixed point is stable and intertangled stochastic patterns of the type discussed in the main paper can develop. 

As a additional  complement, we return on the example of a linear ring made of $\Omega=4$
 nodes, already discussed in the main body of the paper. The PSDM can be readily determined, as discussed in the main article and the associated complex coherent functions computed. 
In left panel of Figure 4 we show $|C_{1,2i+1}|$, with $i = 1, 2, 3$
(plotted on the horizontal axis), for a choice $D < D_c$,
and against $\omega$ (plotted on the vertical axis). The phase
lag, as predicted by the theory, is also displayed in the
Figure: the values reported exactly apply inside the boxes
delimited by the (white) dashed lines. A sequential alternation
of phase and anti-phase synchronization is hence
expected for the stochastic excitatory signals, registered
across the ring. Similar conclusions are drawn when
considering $|C_{1,2i}|$, for with $i = 1, 2, 3$, i.e. the degree
of synchronization between excitators and inhibitors on
different nodes. Patterns of activation instigated
by the noise assisted drive towards self-organization are
hence expected to emerge. Stochastic simulations as reported in the main body of the paper (see also the annexed movie of the dynamics)
confirm the correctness of this conclusion: nodes are termporarily active or inactive, depending
on their position along the chain. The emerging pattern is dynamical and the system 
switches continuously one given configuration and its negative analogue, as time progresses.

\begin{figure}[!h]
 \centering
   {\includegraphics[width=8cm]{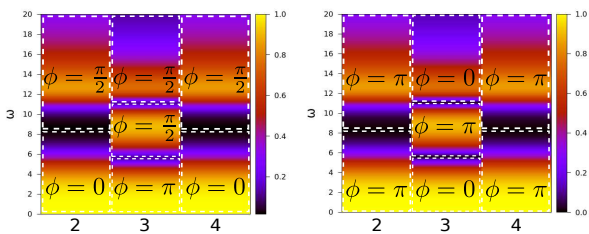}}
   \caption{Left panel: $|C_{1,2i+1}|$ vs. $\omega$ (vertical axis) is plotted, for $i = 1, 2, 3$
(horizontal axis). The system is made up of $4$ nodes organized
in a closed linear ring. The synchronization
occurs for roughly two values of $\omega$, the modulus of the complex
coherence function being more significant at low frequencies.
The phase lag predicted by the theory is also displayed in the
Figure. The values reported apply inside the regions delimited
by the (white) dashed lines Right panel: same for for $|C1,2i|$. }
   \label{fig4}
  \end{figure}


\begin{thebibliography}{99}
\bibitem{footnote1} The model can be complexified as wished, e.g. by enforcing a different fixed point for the two involved species, and the qualitative picture here outlined stays unchanged.
\bibitem{footnote2} Notice that for $D\!=\!0$,  $\lambda_{3,4}$ converge to $\lambda_{1,2}$, as expected.
\end{thebibliography}
 \end{document}